\documentclass[aps,prl,twocolumn,showpacs,superscriptaddress,groupedaddress,graphics]{revtex4}
\usepackage{graphicx}
\usepackage{dcolumn}
\usepackage{bm}
\usepackage{amssymb}
\usepackage{mwe}
\usepackage{comment}

\begin{document}

\title{Maintaining high Q-factor of superconducting YBa$_2$Cu$_3$O$_{7-x}$ microwave cavity\\in a high magnetic field}
\author{Danho Ahn} \affiliation{Center for Axion and Precision Physics Research, Institute for Basic Science, \\Daejeon 34051, Republic of Korea} \affiliation{Department of Physics, Korea Advanced Institute of Science and Technology (KAIST), \\Daejeon 34141, Republic of Korea}
\author{Ohjoon Kwon} \affiliation{Center for Axion and Precision Physics Research, Institute for Basic Science, \\Daejeon 34051, Republic of Korea}
\author{Woohyun Chung} \email{gnuhcw@ibs.re.kr} \affiliation{Center for Axion and Precision Physics Research, Institute for Basic Science, \\Daejeon 34051, Republic of Korea}
\author{Wonjun Jang} \affiliation{Center for Quantum Nanoscience, Institute for Basic Science, \\Seoul 33760, Republic of Korea}
\author{Doyu Lee} \affiliation{Center for Axion and Precision Physics Research, Institute for Basic Science, \\Daejeon 34051, Republic of Korea} \affiliation{Department of Physics, Korea Advanced Institute of Science and Technology (KAIST), \\Daejeon 34141, Republic of Korea}
\author{Jhinhwan Lee} \affiliation{Center for Artificial Low Dimensional Electronic Systems, Institute for Basic Science, \\Pohang 37673, Republic of Korea}
\author{Sung Woo Youn} \affiliation{Center for Axion and Precision Physics Research, Institute for Basic Science, \\Daejeon 34051, Republic of Korea}
\author{Dojun Youm} \affiliation{Department of Physics, Korea Advanced Institute of Science and Technology (KAIST), \\Daejeon 34141, Republic of Korea}
\author{Yannis K. Semertzidis} \affiliation{Center for Axion and Precision Physics Research, Institute for Basic Science, \\Daejeon 34051, Republic of Korea} \affiliation{Department of Physics, Korea Advanced Institute of Science and Technology (KAIST), \\Daejeon 34141, Republic of Korea}
\date{\today}

\begin{abstract}
A high Q-factor microwave resonator in a high magnetic field could be of great use in a wide range of fields, from accelerator design to axion dark matter search. The natural choice of material for the superconducting cavity to be placed in a high field is a high temperature superconductor (HTS) with a high critical field. The deposition, however, of a high-quality, grain-aligned HTS film on a three-dimensional surface is technically challenging. We have fabricated a polygon-shaped resonant cavity with commercial YBa$_2$Cu$_3$O$_{7-x}$ (YBCO) tapes covering the entire inner wall and measured the Q-factor at 4 K at 6.93 GHz as a function of an external DC magnetic field. We demonstrated that the high Q-factor of the superconducting YBCO cavity showed no significant degradation from 1 T up to 8 T. This is the first indication of the possible applications of HTS technology to the research areas requiring a strong magnetic field at high radio frequencies.
\end{abstract}
\maketitle

Superconducting radio-frequency (SRF) science and technology involves the application of superconducting properties in radio frequency systems. Due to the ultra-low electrical resistivity, which allows an RF resonator to obtain an extremely high Q-factor, SRF resonant cavities manifest themselves in a broad scope of application such as quantum devices \cite{SCcavappli_Qubit_01}, material characterization \cite{SCcavappli_MatChar_01, Text_Lancaster, Text_HTSmw} and particle accelerators \cite{SCcavappli_Accel_01, SCcavappli_Accel_02, SCcavappli_Accel_03, SCcavappli_Accel_04, SCcavappli_Accel_05}. However, the presence of an external magnetic field will destroy the superconducting state above the critical field, which limits scientific productivity in many R$\&$D areas such as in high energy particle accelerators \cite{SCcavappli_Accel_03, SCcavappli_Accel_04} and axion dark matter search \cite{AxSearch_Sikivie_SCpaper, AxSearch_Italian_SCpaper}. In particular, the axion dark matter detection scheme utilizes a resonant cavity immersed in a strong magnetic field, by which the axions are converted into microwave photons \cite{AxSearch_Sikivie_AxionSearch_01, AxSearch_Sikivie_AxionSearch_02}. Maintaining high Q-factor of the cavity in the presence of a strong magnetic field at gigahertz (GHz) frequencies is a crucial component in a highly sensitive axion detection system and at the same time, a largely unexplored and challenging task but beneficial to many applications of SRF. The natural choice of material that would be used in fabricating the superconducting cavity under a high magnetic field is the high temperature superconductor (HTS) YBa$_2$Cu$_3$O$_{7-x}$ (YBCO) whose upper critical field is very high and vortex depinning frequency is more than 10 GHz \cite{YBCO_Propert_Vortex_Golosovsky_01, YBCO_Propert_Vortex_Golosovsky_02}. In this work, we preferred to use pure YBCO to other rare-earth barium copper oxide (ReBCO)  materials which have a high concentration of the gadolinium atoms. The RF surface resistance of those ReBCO materials could be higher than that of YBCO because gadolinium is paramagnetic \cite{YBCO_Propert_Vortex_Golosovsky_02, YBCO_Propert_Gadolinium}.\\

It is well known that the Q-factor would degrade with the imperfect grain alignment of YBCO film because the grain boundaries give an additional loss (surface resistance) to the cavity \cite{YBCO_Propert_GrainBoundary}. However, realizing a three-dimensional resonant cavity structure with HTS material whose grains are well aligned, is technically challenging. Directly forming a grain-aligned YBCO film on the deeply concaved inner surface of the cavities is extremely difficult because of the limitation in making the well-textured buffer layers \cite{YBCO_Fabrication_01, YBCO_Fabrication_02, YBCO_Fabrication_03}. Instead, we took advantage of high-grade, commercially available YBCO tapes by AMSC, whose fabrication process, structure, properties are well-known \cite{YBCO_Tape_01, YBCO_Tape_02}. The substrate and stacked buffer layers of the tape were designed to act as template layers to provide the biaxial texture for the YBCO film. The film architecture of the YBCO superconducting tape consists of several parts. On the biaxially textured nickel-tungsten alloy tape, the 800nm thickness YBCO film was deposited on top of buffer layers which consist of Y$_2$O$_3$, YSZ, and CeO$_2$, 75 nm thick each. The YBCO tape is laminated between two stainless steel tapes by soldering to protect the tape from mechanical strain.\\
\begin{figure}
    \centering
    \includegraphics[width=0.48\textwidth]{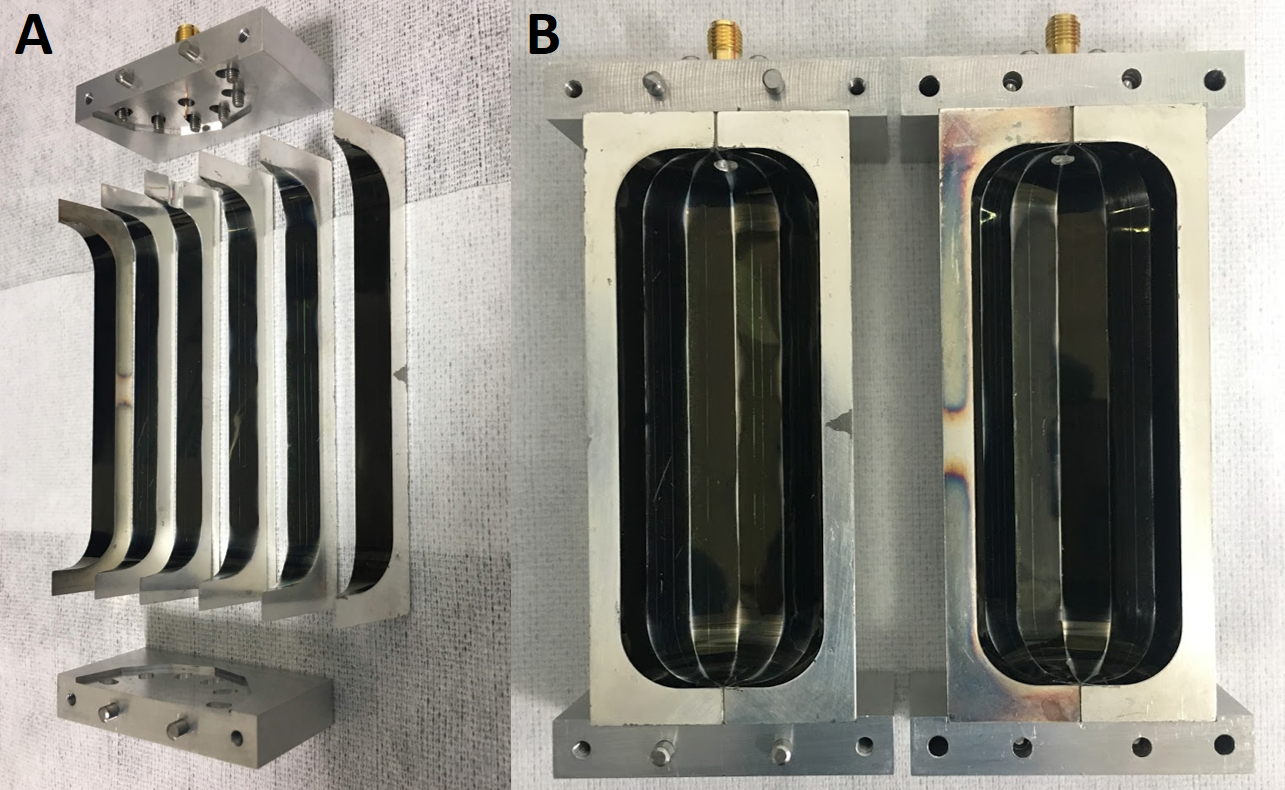}
    \caption{Design of the YBCO polygon cavity. {\bf (A)} Six aluminum cavity pieces to each of which a YBCO tape is attached. {\bf (B)} Twelve pieces composing two cylinder halves are assembled to a whole cavity.}
    \label{fig:01}
\end{figure}
\begin{figure}
    \centering
    \includegraphics[width=0.48\textwidth]{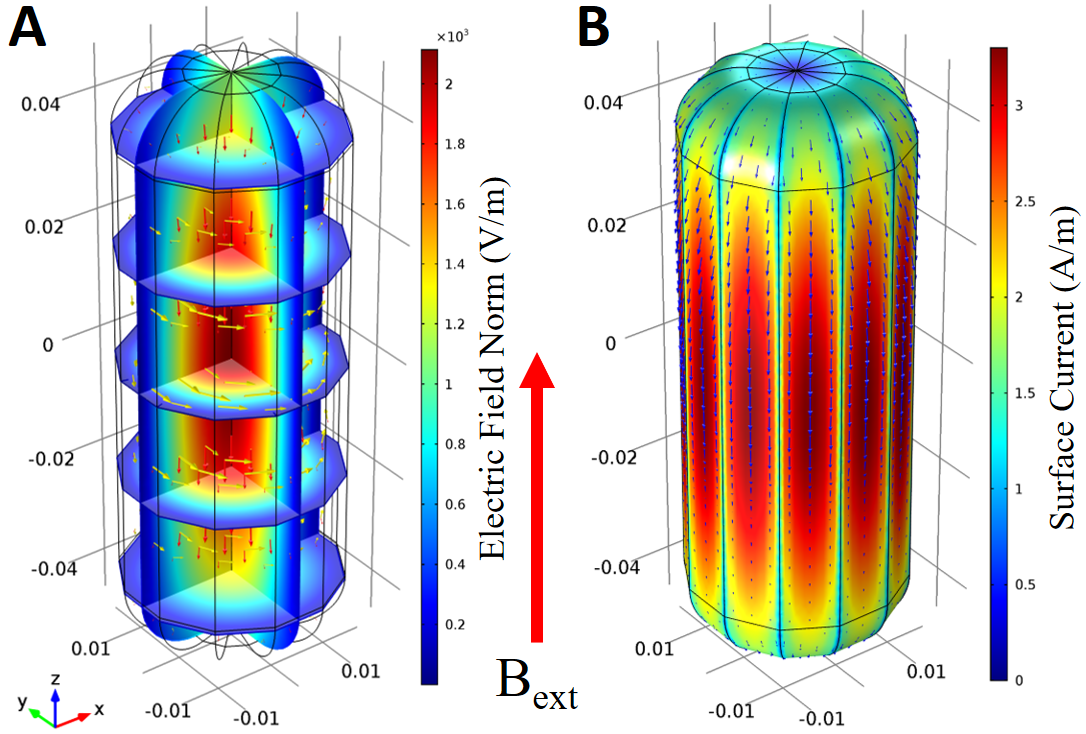}
    \caption{Simulation (COMSOL) result for the TM010 mode in the polygon cavity. B is the direction of the applied magnetic field in the axion cavity experiment. {\bf (A)} The electric field (red arrows, colored 3D plot) and magnetic field (yellow arrows) of the TM010 mode. {\bf (B)} Surface current distribution of the TM$_{010}$ mode. The colored surface shows the magnitude distribution of surface current. Current flows in the direction of the blue arrows.}
    \label{fig:02}
\end{figure}Fabricating a 3-D superconducting cavity utilizing YBCO tapes, we devised an original scheme by employing a 12-piece polygon cavity to which grain-aligned YBCO tapes are attached. Each YBCO tape was prepared and attached securely to the inner surface of a cavity piece with a minimum bending to prevent cracks (FIG.\ref{fig:01}). The arc radius of 10 mm was applied between the top/bottom and the sidewall surfaces to avoid excess bending stress on the YBCO tapes \cite{YBCO_Tape_Bending_01}. The twelve separated cavity pieces are designed for accurate alignment of the YBCO tapes upon assembly.  For the fundamental TM modes most commonly used in axion search, the vertical cuts of the cylindrical cavity do not make any significant degradation of the Q-factor, since the direction of the surface current in TM$_{010}$ mode and the boundary of each cavity piece are parallel as seen in FIG.\ref{fig:02}B, which was already demonstrated by the Center for Axion and Precision Physics research (CAPP) \cite{AxCavity_Vertical_Cut}. We have confirmed it by COMSOL simulation and the Q-factor measurement of an assembled cavity. Once the YBCO tape was completely attached to the inner surface of each polygon piece, we removed the protective layers to expose the bare YBCO surface by a technique developed by CAPP (patent pending). The cut edges of the YBCO tapes exposed on the side were coated by sputtering silver to reduce the loss due to small imperfection created in the cutting process. The technique used in this work was optimized for TM modes of a cylindrical cavity but could be applied to any resonators, minimizing surface losses and resolving contact problems. The assembled cavity was installed in a cryogenic system, equipped with an 8 T superconducting magnet, and brought to a low temperature at around 4 K. The Q-factor and resonant frequency were measured using a network analyzer through transmission signal between a pair of RF antennae, which are weakly coupled to the cavity. The coupling strengths of the antennae were monitored over the course and accounted for in obtaining the unloaded quality factor.\\

\begin{figure}
    \centering
    \includegraphics[width=0.48\textwidth]{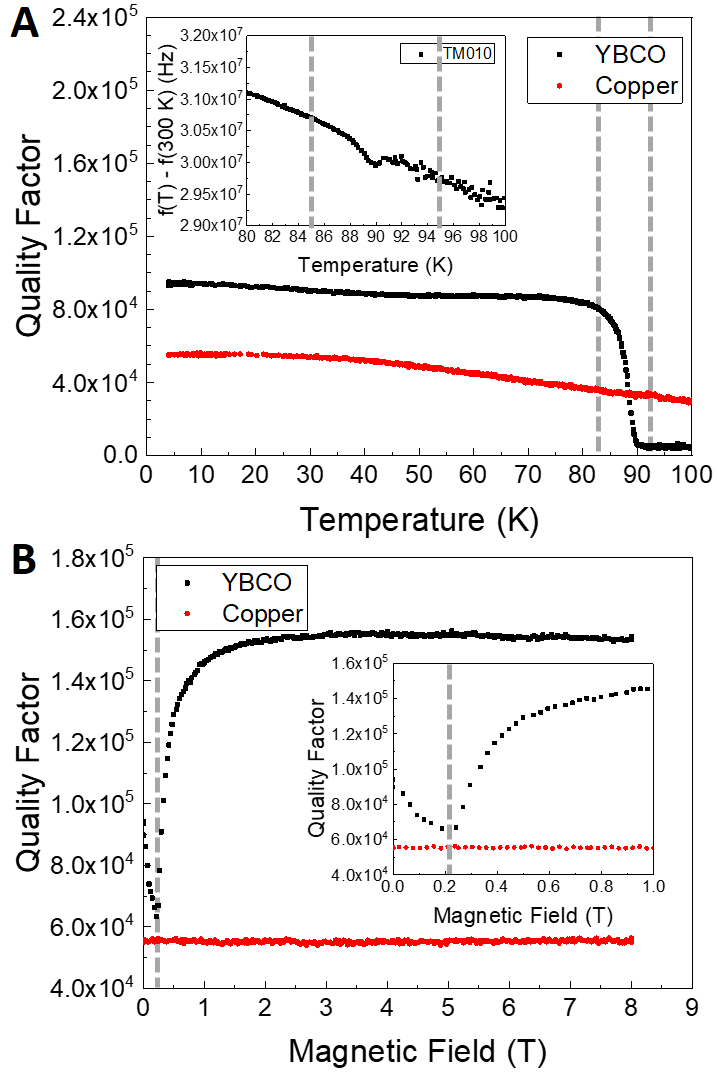}
    \caption{The RF measurement result of the polygon cavities with YBCO (black) and copper (red) inner surface. {\bf (A)} The Q factor measurement data from 4.2 K to 100 K. The inset plot is the frequency shift data ($\Delta$f = f(T) - f(300 K) from 80 K to 100 K. The normal-superconductor phase transition starts at 90 K, at which temperature an anomalous frequency shift occurs. The grey dashed lines show the temperatures 85 K and 95 K. {\bf (B)} The Q factor measurement data from 0 T to 8 T. The dashed line shows the magnetic field 0.23 T at which the abrupt Q factor enhancement starts. The inset plot is the magnified plot from 0 T to 1 T.}
    \label{fig:03}
\end{figure}

Measuring the Q-factor (TM$_{010}$ mode) of the polygon cavity with the twelve YBCO pieces by varying the temperature, we observed the superconducting phase transition around 90 K which agrees with the critical temperature (T$_c$) of the YBCO (FIG.\ref{fig:03}A). The global increase of resonant frequency was due to thermal shrinkage of the aluminum cavity, but an anomalous frequency shift was also observed near the critical temperature. The decrease of the frequency shift at T$_c$ can be attributed to the divergence of the penetration depth of YBCO surface \cite{Text_Lancaster, Text_Tinkham, Text_HTSmw, YBCO_Propert_Vortex_Golosovsky_01}. The maximum Q factor at the 4.2 K was about 95,000. The Q-factor for the same polygon cavity with oxygen-free high thermal conductivity (OFHC) copper (the same geometry) was measured to be 56,500. Varying the applied DC magnetic field from 0 T to 8 T, at the beginning of ramping up the magnet, the Q-factor of the cavity dropped rapidly to ~60,000 until the magnetic field reaches 0.23 T and then rose up to the maximum value of 155,000 at around 3.5 T for the TM$_{010}$ mode. This might be explained in terms of the vibration, movement and the pinning of magnetic vortices, which lies beyond the scope of this work. From the Q-factor measurement, we observed that the Q-factor of the resonant cavity's TM$_{010}$ mode did not decrease significantly (only a few percent changes) until reaching 8 T (FIG.\ref{fig:03}B).\\

The maximum Q-factor achievable with YBCO cavity is currently unknown but the comparison between the surface resistance of 4 K copper (5 m$\Omega$ at 5.712 GHz) \cite{Copper_Rs_01} and 4 K YBCO (0.02 m$\Omega$ at 5-6 GHz) \cite{YBCO_Propert_Vortex_Golosovsky_01, YBCO_DCField_05} suggests that the Q-factor could be much higher even with a strong magnetic field.  Improvements are expected in the near future in techniques of exposing bare YBCO surface from a tape and eventually reducing the area where the surface loss occurs inside the cavity. Our design of the vertically split, polygon-shaped cavity for implementing well-textured YBCO to the inner surface allows us to test the possibility of constructing superconducting resonant cavities which could be used in a strong magnetic field. We demonstrate that it is possible to fabricate a cavity with YBCO inner surface to maintain a high Q-factor even up to 8 T. This result could enable to remove a significant limitation of SRF with a magnetic field in many areas.\\

\section{Acknowledgement}
The authors are grateful for the technical advice of Sergey Uchaikin (Magnetic property of YBCO), Junu Jeong (Data Aquisition).

\end{document}